\documentclass[useAMS,usenatbib]{mn2e}
\usepackage{color}
\usepackage{epsfig}
\usepackage{graphicx}
\usepackage{array}
\usepackage{color}
\usepackage{comment}
\usepackage[utf8]{inputenc}  
\usepackage{amsmath,breqn}  
\usepackage{natbib}
\bibpunct{(}{)}{;}{a}{,}{,}

\usepackage{amssymb}
\usepackage{float}
\usepackage{tabularx}
\usepackage{footnote}
\usepackage{booktabs}
\usepackage[T1]{fontenc}
\usepackage{aas_macros}

\begin{document} 
\title[On the inner disk structure of MWC480]{On the inner disk structure of MWC480: evidence for asymmetries? }
\author[N. Jamialahmadi et al.]{N. Jamialahmadi$^{1,2,3}$\thanks{email: \hspace{0.1cm}  {  jami@ipm.ir  \hspace{0.1cm} \hspace{0.1cm}  and \hspace{0.1cm} jami@oca.eu}},
 B. Lopez$^{3}$,
 Ph. Berio$^{3}$,
  A. Matter$^{3}$,
 S. Flament$^{3}$, 
  H. Fathivavsari$^{1}$,
 \newauthor
 T. Ratzka $^{4}$,
 M. L. Sitko$^{5,6}$,
 A. Spang$^{3}$,
 R. W. Russell$^{7}$\\
 $^{1}$School of Astronomy, Institute for Research in Fundamental Sciences (IPM), P.O. Box 19395-5746, Tehran, Iran \\
 $^{2}$European Southern Observatory, Karl-Schwarzschild. str. 2, D-85748 Garching, Munich, Germany \\
 $^{3}$Laboratoire J.-L. Lagrange, UMR 7293, University of Nice Sophia-Antipolis, CNRS, Observatoire de la Cote D’Azur, \\Boulevard de l’Observatoire - CS 34229 - F 06304 NICE Cedex 4, France.\\
  $^{4}$Institute for Physics / IGAM, NAWI Graz, Karl-Franzens-Universit{\"a}t, 
Universit{\"a}tsplatz 5/II, 8010 Graz, Austria\\
$^{5}$Department of Physics, University of Cincinnati, Cincinnati OH 45221, USA\\
$^{6}$ Space Science Institute, 4750 Walnut Street, Boulder, CO 80303, USA\\
$^{7}$The Aerospace Corporation, Los Angeles, CA 90009, USA\\}


\pagerange{\pageref{firstpage}--\pageref{lastpage}} \pubyear{--}           

\maketitle
\label{firstpage}
\begin{abstract}

Studying the physical conditions structuring the young circumstellar disks is required for understanding the onset of planet formation. Of particular interest is the protoplanetary disk surrounding the Herbig star MWC480. The structure and properties of the circumstellar disk of MWC480 are studied by infrared interferometry and interpreted from a modeling approach.

New observations are driving this study, in particular some recent Very Large Telescope Interferometer (VLTI)/MIDI data acquired in December 2013. 
Our one-component disk model could not reproduce simultaneously all our data: the Spectral Energy Distribution, the near-infrared Keck Interferometer data and the mid-infrared data obtained with the MIDI instrument.
In order to explain all measurements, one possibility is to add an asymmetry in our one-component disk model with the assumption that the structure of the disk of MWC480 has not varied with time. Several senarios are tested, and the one considering the presence of an azimuthal bright feature in the inner component of the disk model provides a better fit of the data.  
\end{abstract}


\begin{table*}
\label{table:1} 
\centering
\begin{tabular}{ c | c | c | c | c}
\hline\hline
 \textbf{Instrument}& \textbf{Date} & \textbf{Telescopes $\&$ Baselines} & \textbf{$B_{p}$ [m]}& \textbf{P.A. [$ ^{\circ} $]}  \\ [0.5ex] 
\hline
MIDI & 2007--02--04&UTs: UT2-UT3&42.8& $52^{\circ }$\\
HIRES&2007--07--03 & Keck1-Keck2 & 84.9 & $48^{\circ }$\\
MIDI&2013--12--29&ATs: K0-J3 & 28.6 & $330^{\circ }$\\[1ex]
 
\hline\hline
\end{tabular}
\caption{Log of long baseline interferometry observations. Our recent MIDI/VLTI observations with ATs (P.I.: Ratzka 2013) were conduced in December 2013. Past observations were made also with the MIDI instrument (P.I.: Di Folco) and with the Keck Interferometer \citep{2009ApJ...692..309E}.}
\end{table*}

\section{\textbf{Introduction}}

Herbig Ae/Be (HAeBe) stars are intermediate mass, pre-main sequence stars (PMS) characterized by the presence of emission lines \citep{2015A&A...579A..81J} and an InfraRed (IR) excess over the stellar photosphere emission. The observed IR excess is caused by circumstellar dust confined into a disk. The study of the structure and physical conditions in such a disk is of great interest for gaining a better understanding of how planetary systems, like our own, are formed. 
         \\

 MWC480, is a Herbig/Ae star (HD 31648, ${A2/3ep+sh}$), with a mass of 1.67$\pm$0.07 $ M_{\odot} $ \citep{2000ApJ...545.1034S}. It is located at d=137$\pm$31 pc \citep{2007A&A...474..653V}. This star is one of the brightest Herbig Ae stars at millimeter wavelengths \citep{1997Natur.388..555M}. The 1.4-mm thermal continuum emission map, obtained with the IRAM telescopes \citep{2000ApJ...545.1034S, 2006A&A...460L..43P} shows a protoplanetary disk with a major axis of 1.20 $ \pm $ 0.15 arcsec (FWHM), corresponding to a half-maximum size $ \sim $ 85 $ \pm $ 20 au. However, the radius of the millimeter dust disk of MWC480 is estimated to be 200 au based on ALMA observations \citep{2017ApJ...835..231H}.
The dusty disk does not show strong large-scale asymmetries, except a putative emission feature extending to the south \citep{2011ApJ...731..133S} and nearly aligned with a jet-like emission observed by \citet{2006AAS...209.3002H}. Large mm grains seem to dominate the sub-mm and mm emission \citep{2011ApJ...731..133S} while radial variations of the dust grain size distribution were also detected with a dust emissivity index increasing with radius \citep{2011A&A...529A.105G}. 
In parallel, the global structure of the gaseous disk, observed in various CO lines, seems to be typical of a disk with continuous surface density distribution evolving by angular momentum transfer through viscous diffusion \citep{2013PASJ...65..123A}. These findings point to a continuous disk with a globally axisymmetric structure at large scale both in gas and dust. \citet{2017ApJ...835..231H} presented ALMA observations at 0.6$\arcsec$ resolution of molecules $DCO^{+} $, $H^{13}CO^{+} $, DCN, and $H^{13}CN$ in a disk with continuous surface density distribution around MWC480. They found that the $H^{13}CO^{+} $ and $DCO^{+}$ radial emission profiles peak at 40 AU and the $H^{+}CN$ profile is centrally peaked. Although the DCN emission was weak, it was consistent with the Keplerian rotation pattern established by the other three lines observed. The DCN emission appeared to feature a central dip, but the signal-to-noise ratio was too low to be definitive. Therefore, the axisymmetric structure were seen in both in gas and dust at ALMA observations.
 
 However, does this apparent axisymmetric and continuous structure translate to smaller scales in the inner disk regions (planet-forming regions)? Can we expect faster and differentiated evolution, with already signposts of disk clearing, in the 0.1-10 au inner region? First hints came from \citet{2008ApJ...678.1070S} who highlighted an IR variability possibly related to a time-dependent shadowing of outer disk areas by the inner disk, knowing that MWC 480 is still actively accreting.
 
Such time-dependent shadowing is supported by a marginal scattered-light detection of the disk by Grady et al. (2010), followed by another detection in scattered polarised light at a time the NIR excess was historically low (Kusakabe et al,. 2012). Since no temperature change was observed at that time, this drop in NIR emission thus very likely correlated with the scale height variability of the inner rim of the dust disk. 
Spatially resolving the inner 0.1-10 au inner disk is thus the key to unveil the origin of this shadowing and more generally the possibility of a differentiated evolution between the inner and outer disk regions. IR stellar interferometry is so far the only technique capable of reaching the angular resolution level of a few milliarcseconds that translates to sub-au scales at the distance of MWC 480.
However, only sparse interferometric data have been obtained on MWC 480. Using the Keck Interferometer, Eisner et al. (2009, 2014) spatially resolved the innermost dust and hot gas disk. In particular, they showed that the Bracket gamma line emission originated mostly inside the dust sublimation radius (< 0.1 au) probably in accretion columns and/or shocks. Moreover, the only MIR interferometric measurement obtained with the VLTI/MIDI was modeled independently by \citep{2015A&A...581A.107M} and \citep{2014Ap&SS.tmp..513J}. Using axisymmetric models, they both showed that most of the MIR continuum emission from warm dust originates within 1-10 au.
Given the very low number of interferometric observations (1 with the VLTI and one with the KI), the putative axisymmetric structure of the inner regions needs to be confirmed with additional measurements obtained along other baseline directions.

 In this context, the aim of this paper is to provide a more detailed description of the inner disk. In particular, we aim to assess the axisymmetric structure of the IR emitting region by combining the existing IR interferometric observations with a new MIDI measurements obtained in 2013 in a perpendicular direction. This new measurement is thus a key to reinforce or not the possibility of an axisymmetric inner disk and unveil signs of on-going physical processes.
 
 Section 2 of this paper summarizes the observations and the data processing. Section 3 shows the results of the
interferometric observations. Section 4 describes our semi-analytical models, and the related results. Section 5 includes a discussion on the modeling results and Section 6
summarizes our work and outlines some observational perspectives to validate the existence of azimuthal asymmetries
in the disks.

 \section{\textbf{Observations}}
   
   \subsection{MIDI observations}

MWC480 was observed in 2007 and in 2013 with the instrument MIDI of the VLTI \citep{2003Ap&SS.286...73L} of the ESO Paranal Observatory. The first observation was carried out on the 4th of February 2007 using two 8 m Unit Telescopes (UTs). The second observation was made on the 29th of December 2013 using two 1.8 m Auxiliary Telescopes (ATs). 
The observations were performed using the prism as dispersive element giving a spectral resolution of R $\sim$ 30 in the N-band for the wavelength range 8--13 $\mu$m using the HIGH-SENS mode. Photometry measurement (total flux) and visibility (normalised coherent flux) were obtained for each observation.
A summary of the observing log, containing the length of the projected baselines is shown in Table 1. 
The UV coverage of all the interferometric observations is shown in Fig.~1 [Top].

 \begin{figure*}
\centering
\begin{tabular}{c}
\includegraphics[clip=,width=0.35\hsize]{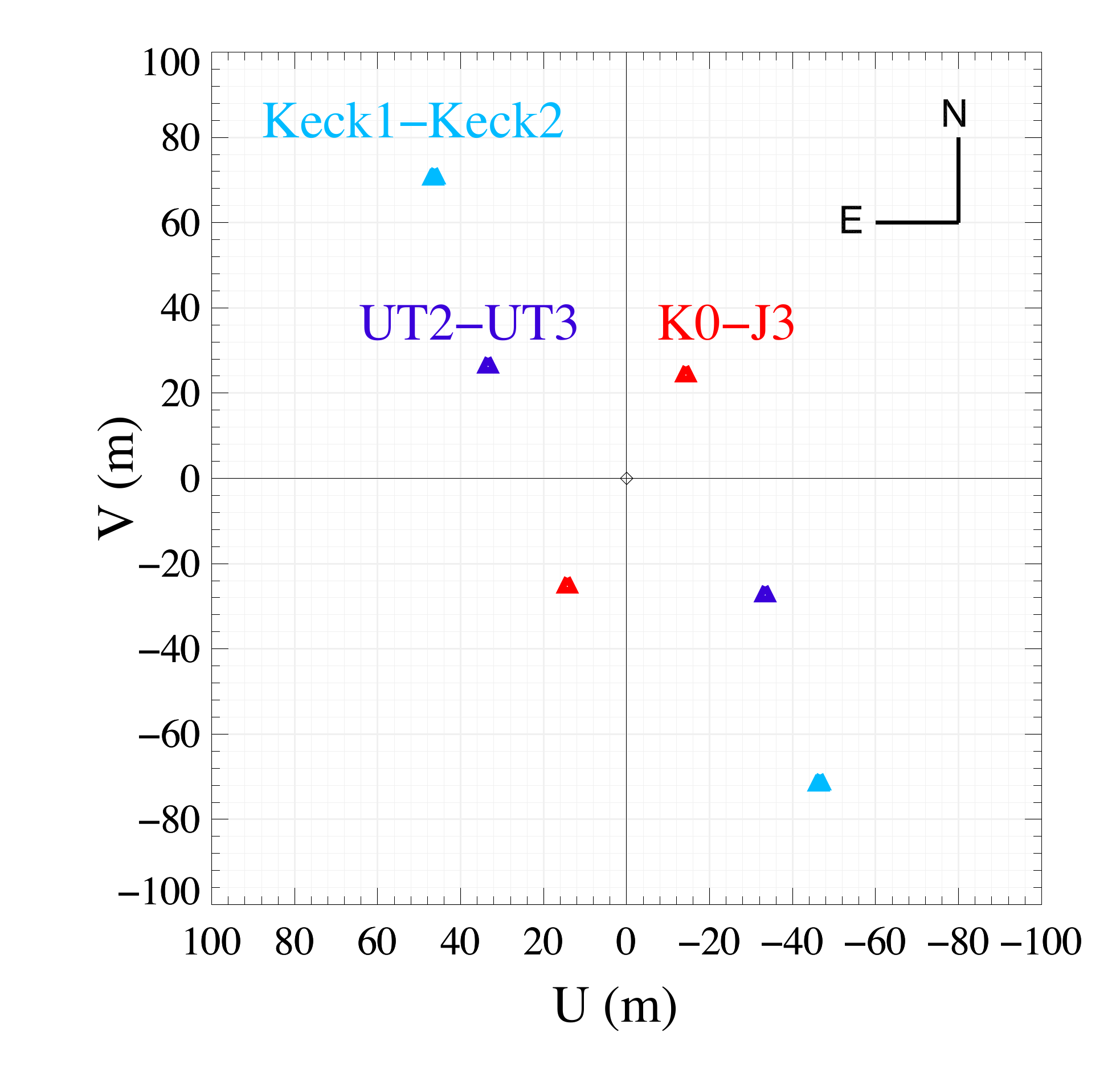} \\
\includegraphics[clip=,width=0.8\hsize]{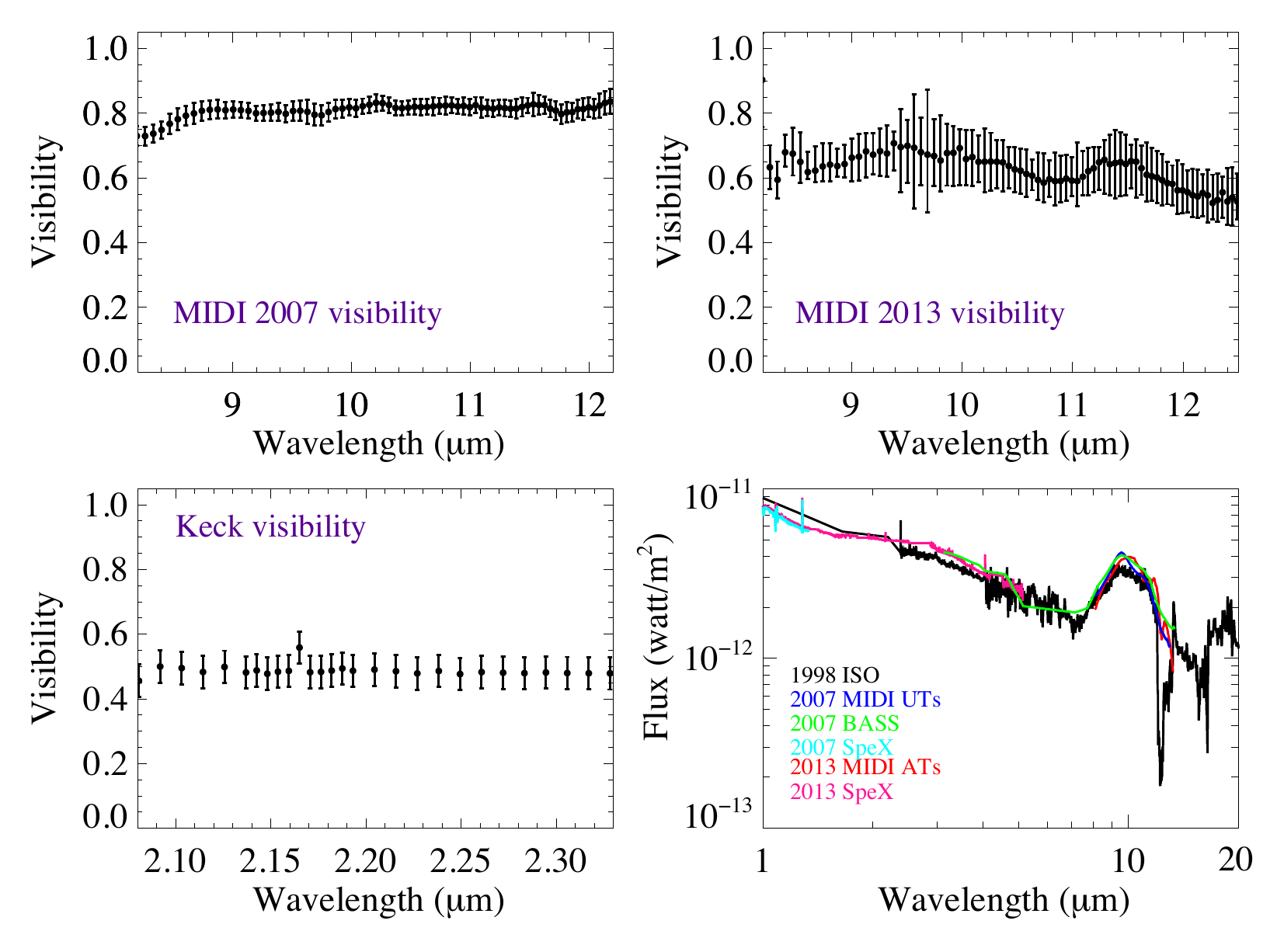}
\end{tabular}
\caption{Top: UV coverage for the three sets Interferometric observations.  Middle-left: Measured N-band visibility of MWC480 as a function of wavelength in 2007 with spectral resolution of R=30. Middle-right: Measured N-band visibility of MWC480 as a function of wavelength in 2013 with spectral resolution of R=30. Bottom-left: Measured K-band visibility of MWC480 (the error bars are represented for few points only). Bottom-right: ISO data for wavelengths 1--20 $ \mu $m in 1998 (black color), SpeX data for wavelengths 1--5.4 $ \mu $m (cyan color) and BASS data for wavelengths 5.5--14 $ \mu $m in 2007 (green color), SpeX data for wavelengths 1--5.4 $ \mu $m in 2013 (pink color), the MIDI total flux for wavelengths 8--13.5 $ \mu $m in 2007 (blue color) and in 2013 (red color).}
 \label{hbalmeroii}
\end{figure*}

The MIDI data reduction was carried out using the EWS (Expert Workstation) software package. 
 EWS performs a coherent analysis of dispersed fringes to estimate the complex visibility of the source. The method and the different processing steps are described in \cite{2004SPIE.5491..715J}. The calibrated visibilities were then obtained by dividing each raw visibility measurement by the instrumental visibility measured on the closest calibrator in time.

In order to calibrate the visibilities of the science source in 2007 and 2013, we used the calibration star HD 20644. This calibrator was selected from the SearchCal tool from the JMMC \footnote{http://www.mariotti.fr/}.

As we can see in Fig. 1 [Middle-left], representing the visibility in 2007, the source is barely resolved (visibility level of about 0.8) and shows a flat visibility profile across the N-band. In contrast, the MIDI visibility obtained in 2013 with a shorter perpendicular baseline, varies from 0.7 to 0.5 across the N-band. This visibility drop stands out in spite of larger error bars due to sensitivity reasons since this visibility was obtained with the smaller ATs.

MWC480 was also observed with the Keck Interferometer in the K band in the wavelengths range 2.08--2.33 $\mu$m \citep{2009ApJ...692..309E}. We downloaded the reduced data from the Keck Archive. According to Fig.1 [Bottom-left], the peak in the data, which has higher visibility, is related to the Br$\gamma$ emission at 2.165 $\mu$m. 

\subsection{Spectroscopic observations with SpeX} 
 We observed MWC 480 with the SpeX spectrograph on IRTF in parallel to the most recent MIDI  observations. The SpeX observations were carried out on September 11th 2013 using the cross-dispersed (hereafter XD) echelle gratings in both short-wavelength mode (SXD) covering 0.8--2.4 $ \mu $m and long-wavelength mode (LXD) covering 2.3-5.4 $ \mu $m \citep{2009ApJS..185..289R}. On nights where the seeing is 1$\arcsec$ or better, this technique yields absolute fluxes that agree with aperture photometry to within 5$\%$ or
better \citep{2015ApJ...805..149I}.  Since our observations were performed with a seeing of 1$\arcsec$, we estimate the SpeX absolute photometric calibration uncertainty to be 5$\%$.

The data were reduced using the Spextool software \citep{2003PASP..115..389V,2004PASP..116..362C}.
Existing Spex data from 2007 by \citet{2007M&PSA..42.5284S} used to cover the epoch of the first MIDI measurement.

According to Fig. 1 [Bottom-right], the 2007 and 2013 MIDI total flux measurements do not present any significant change in amplitude. Moreover, they both appear consistent with the other mid-IR measurements obtained in the same respective epochs, i.e., the SpeX and BASS \footnote{http://irtfweb.ifa.hawaii.edu/Facility/} data. Even the ISO \footnote{http://irsa.ipac.caltech.edu/data/SWS/}  spectrum obtained in 1998 appeared very similar. 

Between 2007 and 2013, the NIR and MIR variability do not appear anyway more than 10$\%$. Interestingly, this relative error is what we estimated for the absolute calibration error of our MIDI data and is greater than what we estimated for the absolute calibration error of our SpeX data.

\section{\textbf{Modeling}}

To interpret in a consistent way all our measurements, i.e., the SED, the Keck visibility and the MIDI visibilities, we developed a semi-analytical model. 
 
Our model is based on the hypothesis of a temperature and surface density-gradient for the circumstellar disk comparable to what was used for previous disk descriptions \citep[e.g,][]{1998ApJ...492..540H}. Such a disk model is defined by an inner radius $r_{\rm in}$ and an outer radius $r_{\rm out}$, with temperature and surface density profiles that are parameterised by power laws: 
 
\begin{equation}
T_{r} =T_{\rm in}\left(\frac{r}{r_{\rm in}}\right)^{-q}, 
\end{equation} 
 
\begin{equation}
\Sigma_{\rm r} =\Sigma_{\rm in}\left(\frac{r}{r_{\rm in}}\right)^{-p},
\end{equation}
 
 with q ranging from 0.5 (flared irradiated disks) to 0.75 (standard viscous disk or flat irradiated disks), see e.g., \citet{1981ARA&A..19..137P}.
 $T_{in}$ is the temperature of a free grain\footnote{A grain illuminated directly by the central star assuming no radiative exchange with other grains.} located at $r$=$r_{in}$ which is the inner radius of the disk. 
 
 Equating the absorption and the emission of a grain with non-chromatic (grey) absorptivity,  one can calculate the $T_{in}$ at $r_{in}$ in Eq.(1):
   \begin{equation}
      T_{in}=T_ {\star} \left(\frac{R_{\star}}{2r_{in}}\right)^{\frac{1}{2}}\,,
   \end{equation}
    where $T_ {\star}$ (=\,8970\,K) and $R_{\star}$ (=\,0.068\,mas) are the stellar effective temperature and the stellar radius and their respectively.
    
    In the models of \citet{2001ApJ...560..957D} and \citet{1997ApJ...490..368C}, $\Sigma_ {in}$ has been considered at r=1 AU, while in e.g., \citet{1998A&A...338L..63D}, \citet{2011A&A...535A.104D}, who used milimeter observations, $\Sigma_ {\rm in}$ is assumed to be defined from the outer radius of the disk. $\Sigma_ {\rm in}$ is related to the total mass amount of the dust. The mass of the dust is given by
  \begin{equation}
       M_{dust}=\int_0^{2\pi} \int_{r_{in}}^{r_{out}} \mathrm\Sigma_{\rm r} r \,\mathrm{d}r \mathrm{d}\theta\
   \end{equation} 
  Combining Eq.(2) and Eq.(4) gives
  \begin{equation}
       \Sigma_{\rm in}=\frac{M_{dust}}{2\pi {r}_{in}^{p} f}\,,
   \end{equation} 
   where 
   \begin{equation}
   f=\frac{1}{2-p} \left[\left(\frac{r_{out}}{r_{in}}\right)^{2-p}-1\right]
   \end{equation} 

          p value in the Eq.(2) varies in different studies. Isella et al. (2009) show that p ranges from -0.8 to 0.8 based on observations of low and intermediate mass pre-main sequence stars. Assuming constant mass accretion rate at constant viscosity p value is 1.  p is assumed to be 1.5 for the MMSN (Minimum Mass Solar Nebula) \citep{1997LPI....28.1517W} and assumed often as a basis in other disk models (e.g., \citet{1997ApJ...490..368C}; \citet{2001ApJ...560..957D}; \citet{2009ApJ...692..309E}). p is assumed not to be 2 in our models.
\begin{figure*}
 \centering 
      \includegraphics[width=16cm, height=6cm]{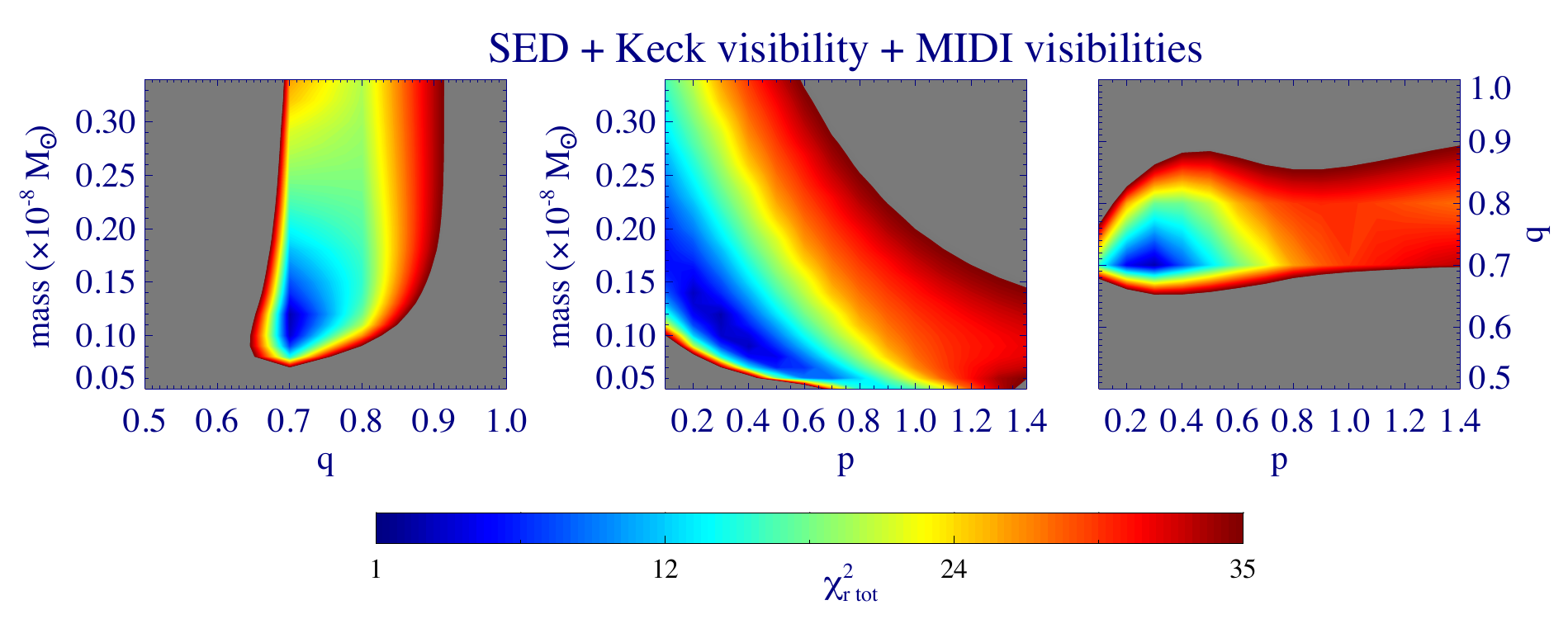}
   \caption{Minimum ${\chi^{2}} $maps for the SED, the Keck visibility and the MIDI visibilities for the one-component disk model. We show the minimum ${\chi^{2}} $ for each pair of parameters. The grey areas are related to the ${\chi^{2}} $ larger than 35.}
  \label{fig:3}
 \end{figure*}

\begin{table*}
\label{table:3}  
\setlength{\tabcolsep}{3.5pt}   
\centering
\begin{tabular}{c c c c c}
\hline
\hline
Scan range & p & q & $M_{\rm dust}$& $n_{\rm steps}$\\ 
              &   &   &  [${M_{\odot}}$]&\\ 
\hline
Wide& [0.1-1.5] &[0.1-1.5]& [0.01-0.5] $\times$  $10^{-8}$ & 30   \\
Narrow& [0.1-0.6] &[0.5-0.9]& [0.05-0.2] $\times$  $10^{-8}$ & 15  \\
Best-fit& 0.3$\pm$0.08& 0.7$\pm$0.02&(0.11$\pm$0.03)$\times$  $10^{-8}$ \\
\hline
\end{tabular}
\caption{Scanned parameter range of one-component disk model and the best-fit values for each parameter. The 1-$\sigma$ uncertainties on the parameters have been shown as well.  }   
\end{table*}

In our disk model, the observer receives for each disk elementary surface area A\footnote{The elementary surface area of the disk is defined by our pixel size in the simulated brightness maps of our model.}:   
\begin{equation}
       dF_{\lambda}(i)=B_{\lambda} \left[  T(r)\right] \left[1-exp\left(-\frac{\tau_{\lambda}(r)}{cos(i)}\right)\right]  \left(\frac{A}{D^{2}}\right),
   \end{equation} 
where, $i$ is the disk inclination and the quantity $A/{D^{2}}$ represents the solid angle of each elementary surface area as seen at the distance $D$. $\tau_{\lambda}(r)=\Sigma(r)\times\kappa_{\lambda}$ represents the optical depth in the vertical direction, with $\kappa_{\lambda}$ the mass absorption coefficient. For the latter, we used $\kappa_{\lambda}$ from \citet{2011MNRAS.412..711T}, which is computed from Mie theory. They considered a power law (proportional to $a^{-3.5}$) for the grain size distribution with a minimum size of  $a_{max}$=0.02 $\mu$m and three values for the maximum size of $a_{max}$=[10, 50, 200] $\mu$m.\\
The interstellar dust optical constants of \citet{1993ApJ...402..441L} for amorphous grains are considered.

The stellar contribution is represented in our model by $B_{\lambda}( T_{\star} )$ $\times$ ($\pi$ ${\alpha_{\star}}^{2}$). Here $\alpha_{\star}$=$R_{\star}/D$ is the stellar angular radius.\\ 
To take into account the Br$\gamma$ emission  line present in our K band data, we modelled circumstellar Br$\gamma$ emission at 2.165 $\mu$m by including a narrow optically thin isothermal gaseous disk in the inner region of the disk. We based our modelling of the Br$\gamma$  emission on the study done by \citet{2009ApJ...692..309E}.

 \begin{figure*}
\centering
\begin{tabular}{c}
\includegraphics[clip=,width=0.8\hsize]{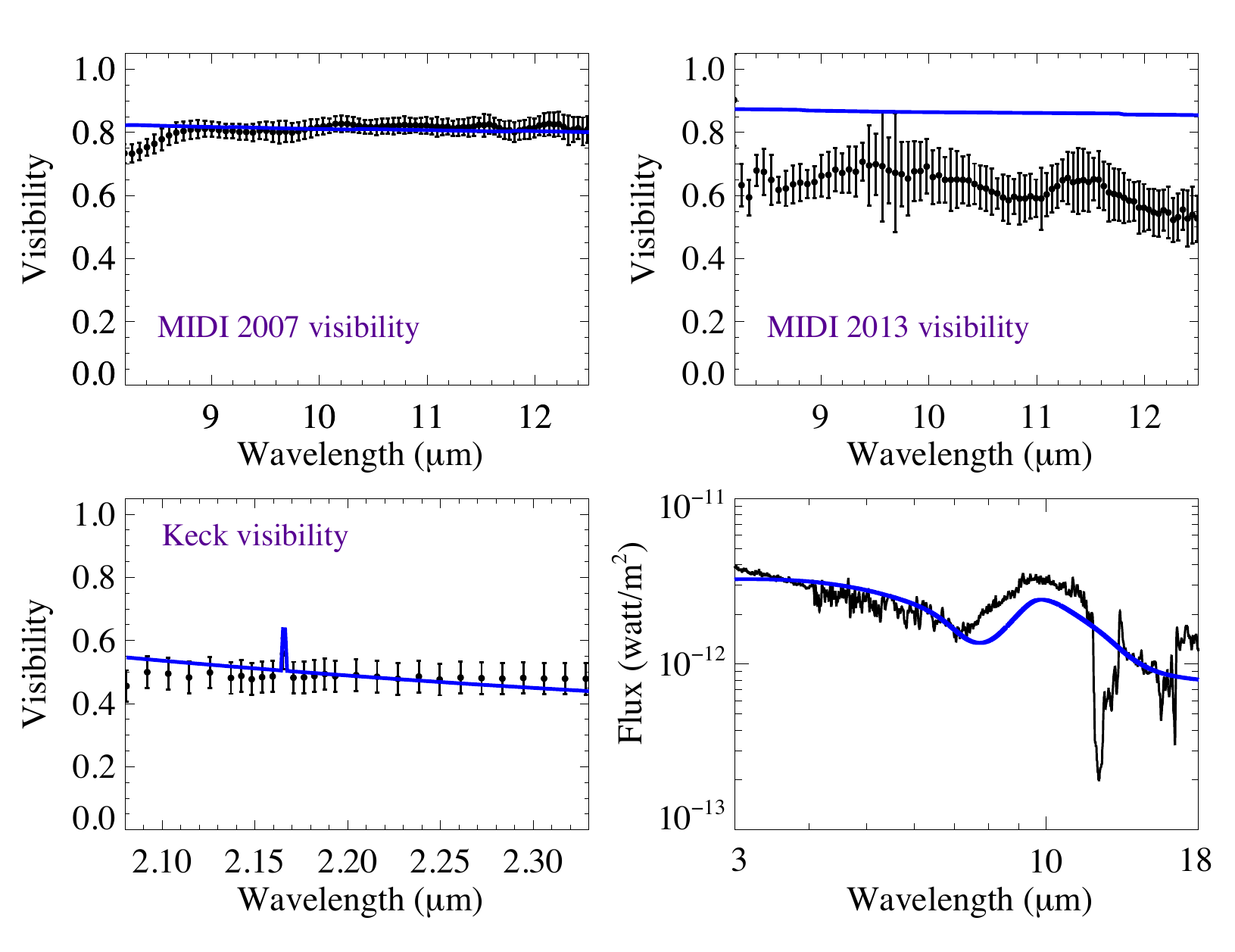} \\
\includegraphics[clip=,width=0.4\hsize]{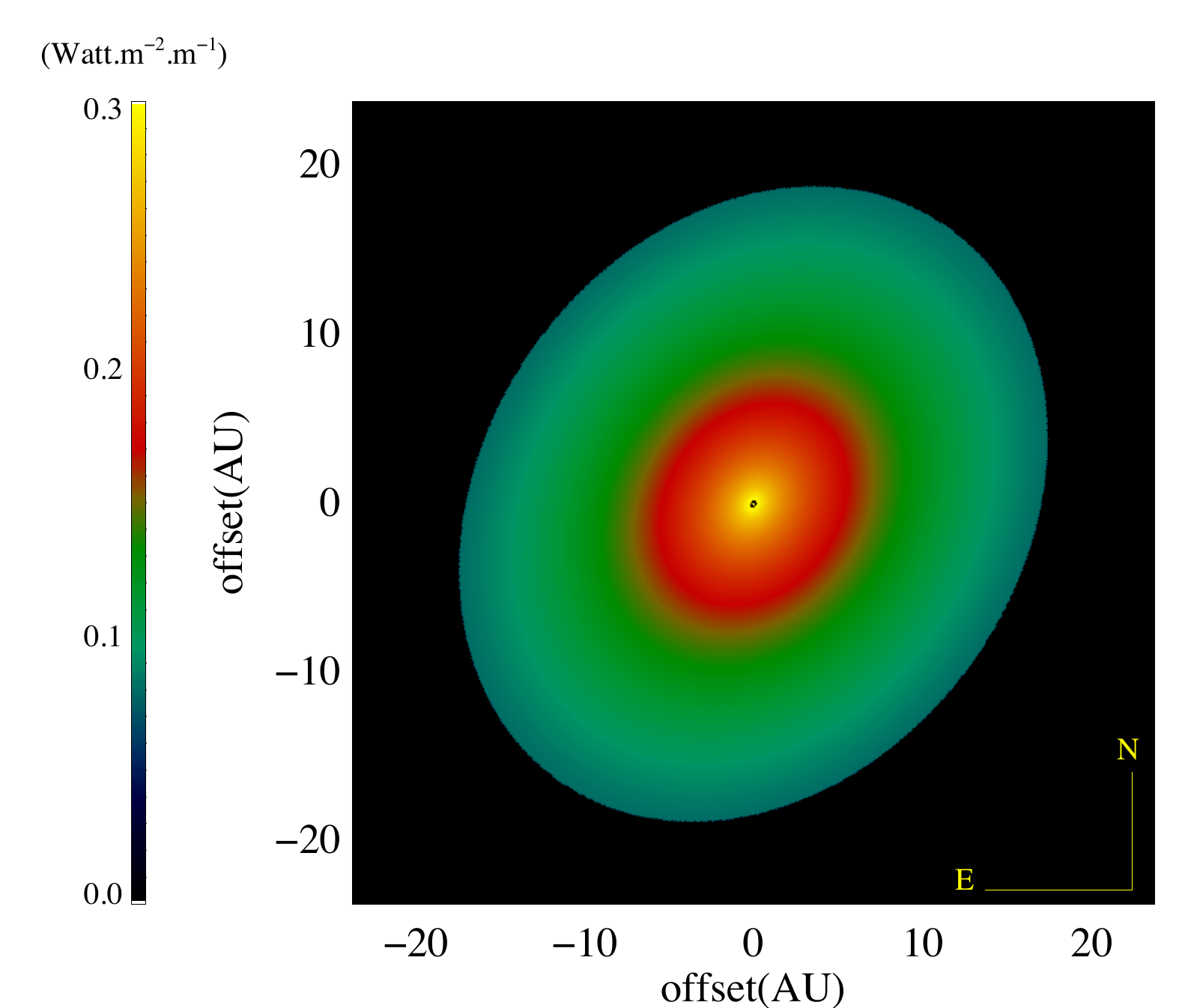}
\end{tabular}
\caption{One-component disk model. Top: The MIDI visibilities in 2007 and 2013. Middle-left: The Keck visibility. Middle-right: The SED from 3 $\mu$m to 18 $\mu$m. The black lines in whole plots represent observations and the blue ones represent the best model. Bottom: The synthetic image of MWC480 at $\lambda$=10 $\mu$m.}
 \label{hbalmeroii}
\end{figure*}
 
After creating the image of the disk, which is the sum of the flux of the central star and the disk for each disk elementary surface area, one can calculate the visibility for each wavelength. We do the fourier transform of the image at each wavelength and the visibility is the modulus of the fourier transform divided by the total flux, which is equal to the fourier transform at zero spatial frequency.


\subsection{Axisymmetric model}

As a first step, we model the inner disk of MWC480 with an axisymmetric one-component disk with a continuous radial structure.  \citet{2007ApJ...669.1072E} already tried to model the IR excess of this source assuming an optically thick disk emission. However, such a model could not account for all the IR excess between 2 and 10 $\mu$m. In our case, we rather use a disk model that includes radial temperature and surface density profiles (see. Eq1 and Eq.2). Our model thus takes into account the disk vertical optical depth towards the observer. 
To limit the number of free parameters, the inner radius is fixed to 0.27 au (Eisner et al. 2009). Since our study focusses on the NIR and MIR emission, which comes mainly from the 0.1 to 10 au of the disk \citep{2015A&A...581A.107M}, we set the outer radius of the disk to 20 au. Reproducing longer wavelength emission is out of the scope of this article. The inclination and the position angle of the disk are also fixed to i=$37^{\circ}$ and PA=$140^{\circ}$ respectively (Chapillon et al. 2012) and (Pietu et al. 2007). As we mentioned in section 3, we adopted the grain size distribution used by \citet{2011MNRAS.412..711T} with a minimum size of $a_{max}$=0.02 $\mu$m and a maximum size of $a_{max}$=10 $\mu$m.

The free parameters of the one-component disk model are thus:
\begin{enumerate}
\item[-] the temperature power law exponent  q,
\item[-] the surface density power law exponent  p,
\item[-] the dust mass $ M_{dust} $, 
\end{enumerate}

In order to obtain the best-fit model, first we scanned a wide range of values for each parameter and minimized the global $ \chi^{2} $ between the model and the observations. The $ \chi^{2} $ maps were obtained for the SED, the MIDI and the Keck visibilities as following:


\begin{equation}
 {\chi^{2}}_{\rm SED}  =  \sum_{\rm i=1}^{N_{\rm SED}}         \frac{({F_{\rm model}}(\lambda_{\rm i})-{F_{\rm obs}}(\lambda_{\rm i}))^{2}}   {{\sigma^{2}}_{{F_{\rm obs}}(\lambda_{\rm i})}}  
\end{equation}

\begin{equation}
 {\chi^{2}}_{\rm vis1}  =  \sum_{\rm k=1}^{n_{\rm baseline}}    \sum_{\rm j=0}^{N_{\rm visMIDI}}         \frac{({V_{\rm model}}(\lambda_{\rm j},B_{\rm k})-{V_{\rm MIDI}}(\lambda_{\rm j},B_{\rm k}))^{2}}   {{\sigma^{2}}_{{vis}(\lambda_{j})}}  
\end{equation}

\begin{equation}
 {\chi^{2}}_{\rm vis2}  =   \sum_{\rm l=0}^{N_{\rm visKeck}}         \frac{({V_{\rm model}}(\lambda_{\rm l})-{V_{\rm Keck}}(\lambda_{\rm l}))^{2}}   {{\sigma^{2}}_{vis(\lambda_{\rm l})}}  
\end{equation}

\begin{equation}
 {\chi^{2}}_{\rm tot}  = {\chi^{2}}_{\rm SED} +{\chi^{2}}_{\rm vis1} + {\chi^{2}}_{\rm vis2}    
\end{equation}

\begin{equation}
 {\chi^{2}}_{\rm r \hspace{2pt} \rm tot}  = \frac {{\chi^{2}}_{\rm tot} } {(N_{\rm SED} + N_{\rm visMIDI} \times n_{\rm baseline} + N_{\rm visKeck} -3)}  
\end{equation}

\begin{table*}
\label{table:3}  
\setlength{\tabcolsep}{3.5pt}   
\centering
\begin{tabular}{c c c c c}
\hline
\hline
Scan range & p & q & $M_{dust}$&$n_{\rm steps}$\\ 
              &   &   &  [${M_{\odot}}$]\\ 
\hline
Wide& [0.1-1.5] &[0.1-1.5]& [0.01-0.5] $\times$  $10^{-8}$& 30   \\
Narrow& [0.4-0.8] &[0.2-0.7]& [0.008-0.04] $\times$  $10^{-8}$ & 15  \\
Best-fit& 0.6$\pm$0.02& 0.5$\pm$0.03&(0.01$\pm$0.006)$\times$  $10^{-8}$ \\
\hline
\end{tabular}
\caption{Scanned parameter range of one-component disk model for the best fit of MIDI visibility in 2013 and the best-fit values for each parameter. The 1-$\sigma$ uncertainties on the parameters have been shown as well.  }   
\end{table*}

 \begin{figure*}
\centering
\begin{tabular}{c}
\includegraphics[clip=,width=0.8\hsize]{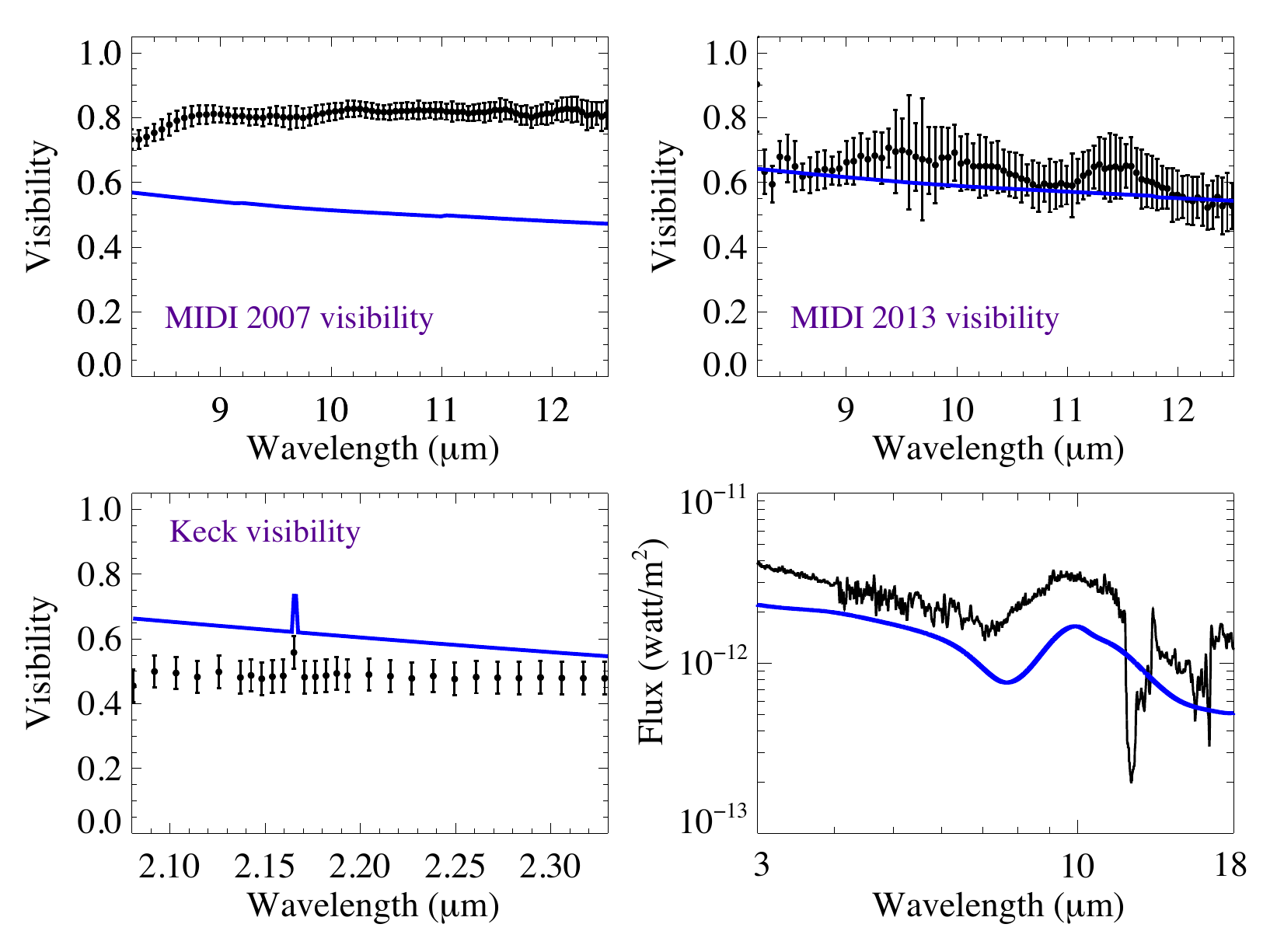} 
\end{tabular}
\caption{One-component disk model for the best-fit of MIDI visibility in 2013. Top: The MIDI visibilities in 2007 and 2013. Bottom-left: The Keck visibility. Bottom-right: The SED from 3 $\mu$m to 18 $\mu$m. The black lines in whole plots represent observations and the blue ones represent the best model}
 \label{hbalmeroii}
\end{figure*}

 For the SED fitting, we used ISO data in 1998 and considered the wavelength range between 3 $\mu$m to 13 $\mu$m. For the MIDI visibilities, we considered the wavelength range between 8.2 $\mu$m to 12.5 $\mu$m. For the Keck visibility, we considered the wavelength range between 2.08 $\mu$m to 2.33 $\mu$m. Fig. 2 illustrates the $ \chi^{2} $ maps for the best parameters. A minimum $ \chi^{2} $ can be identified in the maps. The value of the ${\chi^{2}}_{\rm r \hspace{2pt} \rm tot} $ for the wide range of parameters was 1.38.
In a second step, we scanned on a narrower range around the global minimum $ \chi^{2} $ to refine the best-fit model and the estimation of the best-fit parameters. Table 2 shows the wide and narrow value ranges that were scanned for every parameter. The uncertainties of the best parameters were derived using a Monte Carlo procedure.  To this aim, assuming a normal error distribution, 1000 random data sets corresponding to the measured values varied among their 1-$\sigma$ uncertainties were simulated.  In the next step, calculating the standard deviation of each best-fit parameter value matching to the simulated data sets, one could derive the uncertainties on the parameters (see Table 2).

 \begin{figure*}
\centering
\begin{tabular}{c}
\includegraphics[clip=,width=0.8\hsize]{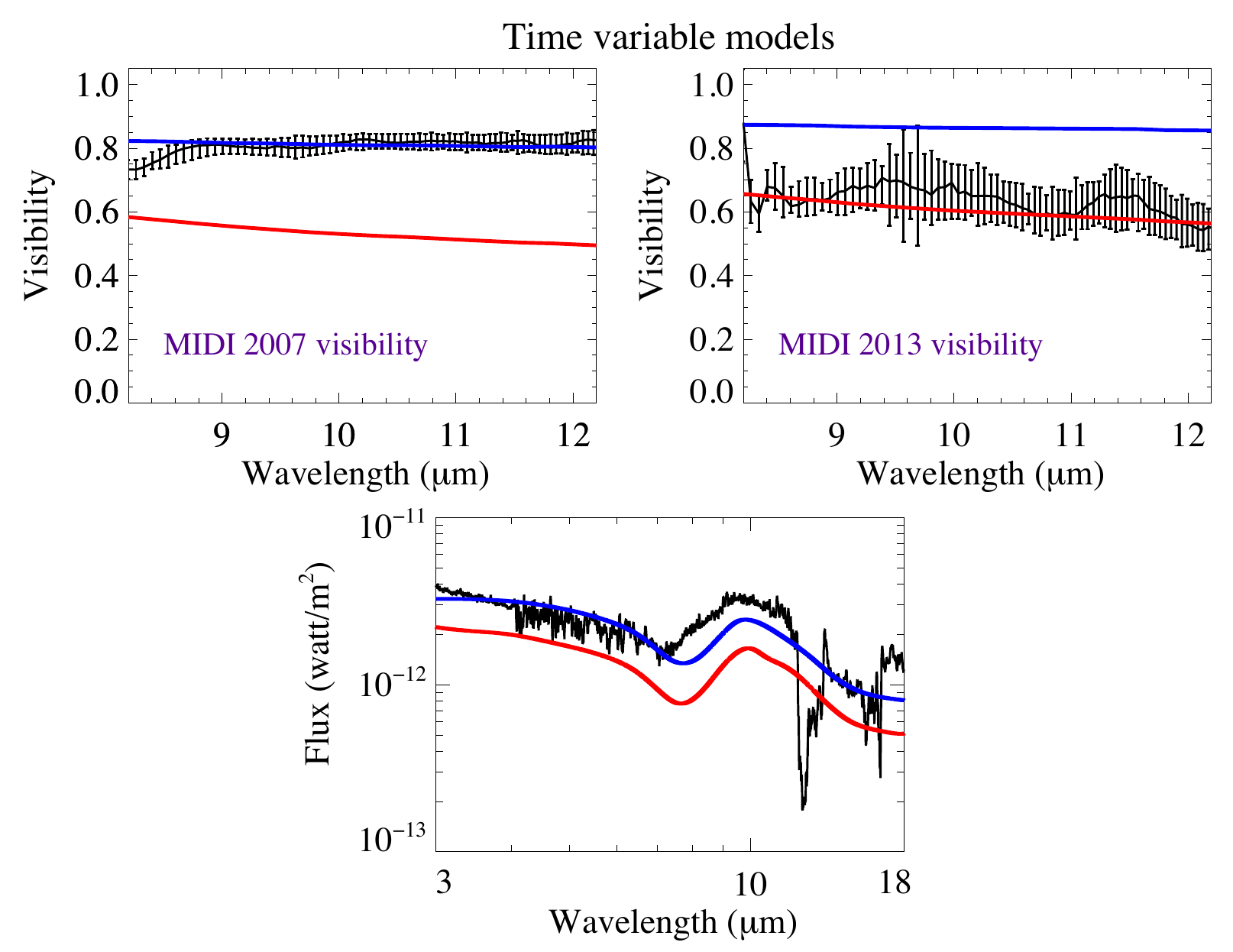} 
\end{tabular}
\caption{Time variable models in 2007 and 2013. The blue solid lines are related to the best-fit model,which reproduces the SED and the MIDI visibility in 2007 for the maximum grain size of 10 $\mu$m. The red solid lines are related to the model, which reproduces the MIDI visibility in 2013 for the maximum grain size of 10 $\mu$m. }
 \label{hbalmeroii}
\end{figure*}

Using the best-fit parameter values we derived for the one-component disk model, we plot our measurements in Fig. 3. The SED mostly up to 8 $\mu$m, the Keck visibility and the 2007 MIDI visibility are reasonably well reproduced, as already shown by \citet{2014Ap&SS.tmp..513J}. However, our best-fit model cannot explain well the 2013 MIDI visibility and the SED between 8 to 13 $\mu$m.  We obtained this best-fit model using a maximum grain size of 10 $\mu$m, which is consistent with the silicate grain emission features. We plot the synthetic image of MWC480 at $\lambda$=10 $\mu$m (Fig. 3-bottom). \\

 \begin{figure}
 \centering 
      \includegraphics[width=7cm, height=5cm]{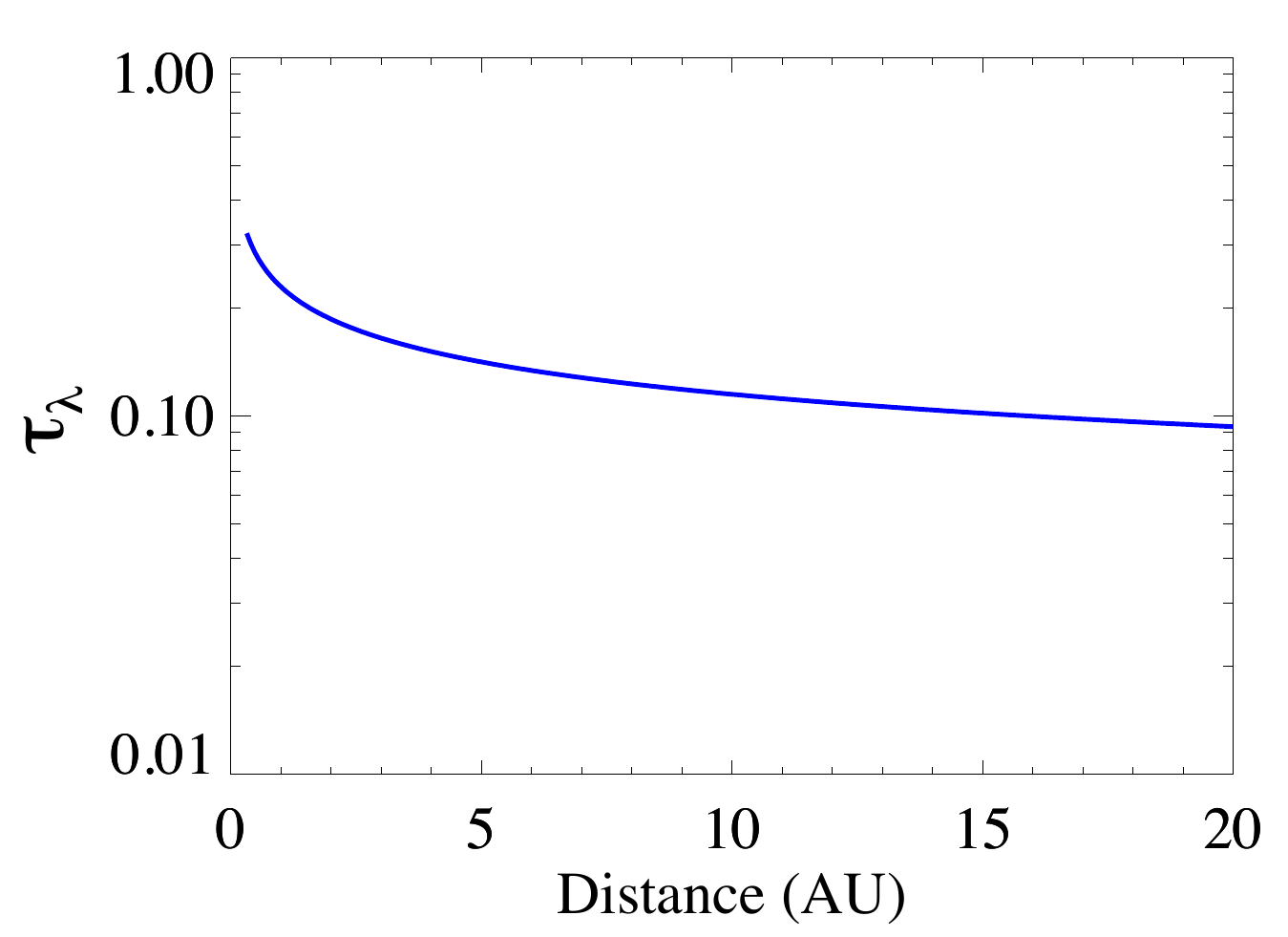}
   \caption{The vertical optical depth versus the distance from the central star.}
  \label{fig:3}
 \end{figure}

 \begin{figure*}
 \centering 
      \includegraphics[width=16cm, height=6cm]{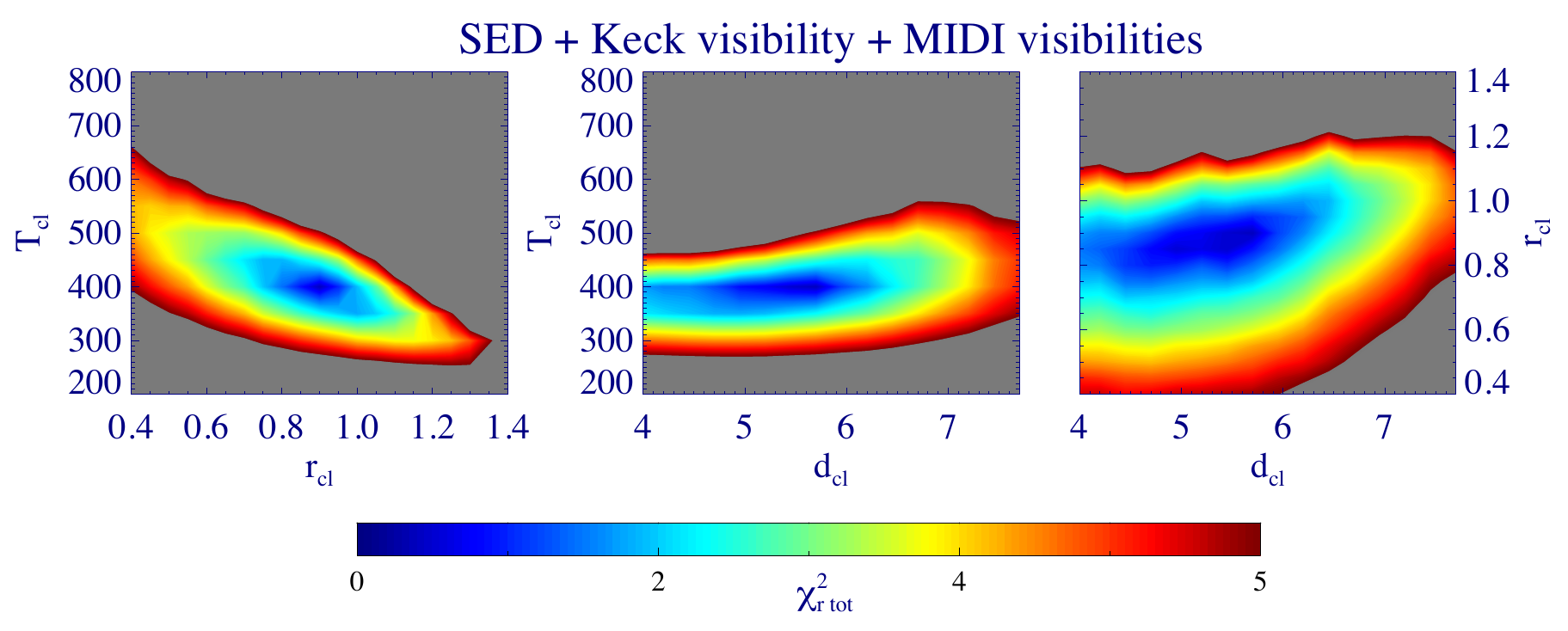}
   \caption{Minimum ${\chi^{2}} $maps for the SED, the Keck visibility and the MIDI visibilities. We show the minimum ${\chi^{2}} $ for each pair of parameters for the one-component disk model with the clump. The grey areas are related to the ${\chi^{2}} $ larger than 5.}
  \label{fig:3}
 \end{figure*}

As a caveat, we remind that Eq.7 does not include a distinct term for the continuum emission, which implies that our model aims to reproduce both the IR continuum and the silicate emission band.
As a consequence, $M_{dust}$ should be seen as a scaling factor for the IR emission level rather than a pure estimation of the actual mass of dust in the inner disk. Moreover, the corresponding vertical optical depth is always less than 1 (see Fig. 6). Therefore, the disk is far from being optically thick in the IR. If one is interested in adding Far-IR emissions to the near and mid-IR regions, indeed in that case, one needs to model the disk midplane where larger grains accommodate and are responsible for the far-IR emissions. Therefore, in that case we may have better estimation of the mass of the disk. However, we remind that the scope of this paper is to model the IR emission coming from the inner disk.  

\begin{table*}
\label{table:4}  
\centering
\begin{tabular}{c c c c}
\hline
\hline
Scan range & $ d_{\rm cl} $ & $r_{\rm cl} $ & $T_{\rm cl} $\\ 
             &  au &  au &  K\\  
 \hline
 Wide& [1-10] &[0.1-2]& [100-1000]   \\
Narrow& [4-7] &[0.6-1.2]& [300-500]   \\
Best-fit& 5.45$\pm$0.8& 0.85$\pm$0.12&400$\pm$57 \\
\hline
\end{tabular} 
\caption{Scanned parameter range of the clump in the one-component disk model and the best-fit values for each parameter. The 1-$\sigma$ uncertainties on the parameters have been shown as well.} 
\end{table*}

\subsubsection{Time variable models} 

As mentioned in section. 2.2, the mid-infrared spectra taken at two epochs in 2007 and 2013 are consistent with each other within their uncertainties. This fact gives us a possibility to assume that the structure of the disk of MWC480 has not varied in 2007 and 2013. However, in order to exclude the temporal variability as the source of changes in the visibilities in these two years one should demonstrate that fitting independently the two MIDI epochs would produce a photometric variability that is inconsistent with the observed mid-infrared spectra, which is dominated by the silicate emission. To this aim, we tried to explore the free parameters of p, q and $M_{dust}$ to find the best model, which can fit the MIDI visibility in 2013. We followed the same method used in Section. 3.1 to calculate the $\chi^{2}$  fitting process and the uncertainties of the parameters. Contributing the SED, the Keck and the MIDI visibility in 2007 in the $\chi^{2}$ fitting process was not effective to obtain the best-fit MIDI visibility in 2013. Consequently, we included only the MIDI visibility in 2013 in the $\chi^{2}$ fitting process to obtain the best-fit parameters values. Table. 3 shows the best-fit parameters and their uncertainties. Fig. 4 illustrates the best-fit model, which can fit only the MIDI visibility in 2013 but not the SED, the Keck and the MIDI visibility in 2007 simultaneously.

We then did a comparison between our best-fit time variable models that are consistent with the 2007 and the 2013 MIDI visibilities individually.
In Fig. 5 we illustrate the time-variable models in 2007 and 2013 for the SED and the MIDI visibilities. The blue solid lines show the best-fit model consistent with the MIDI visibility in 2007. In this model, we could reproduce the 2007 MIDI visibility and the SED but not the MIDI visibility in 2013 at the same time. The red solid lines show the model, which can explain the 2013 MIDI visibility but not the SED and the MIDI visibility in 2007. Indeed, our axisymmetric time-variable models produce the photometric variability that is inconsistent with the observed mid-infrared spectra. This means that our assumption that the structure of the disk of MWC480 has not varied in 2007 and in 2013 based on their consistent MIDI spectra can be truly considered. 

{

As a conclusion, an axisymmetric model, time-variable model or not, appears not to be anymore consistent with our data when taking into account the 2013 MIDI visibility. In the next section, we explore the possibility of an asymmetric disk model to reproduce in a more consistent way our dataset.

  \begin{figure*}
\centering
\begin{tabular}{c}
\includegraphics[clip=,width=0.8\hsize]{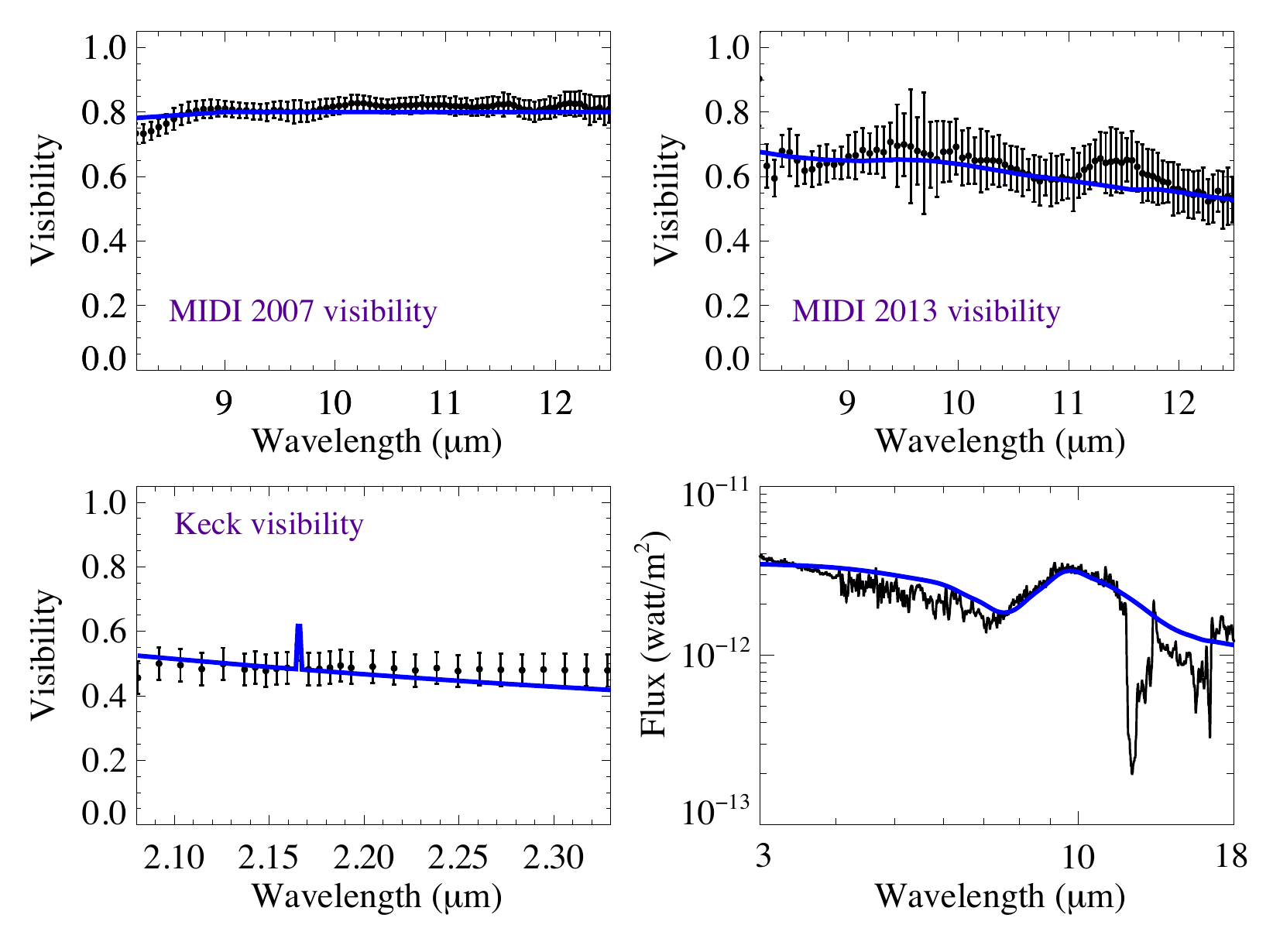} \\
\includegraphics[clip=,width=0.4\hsize]{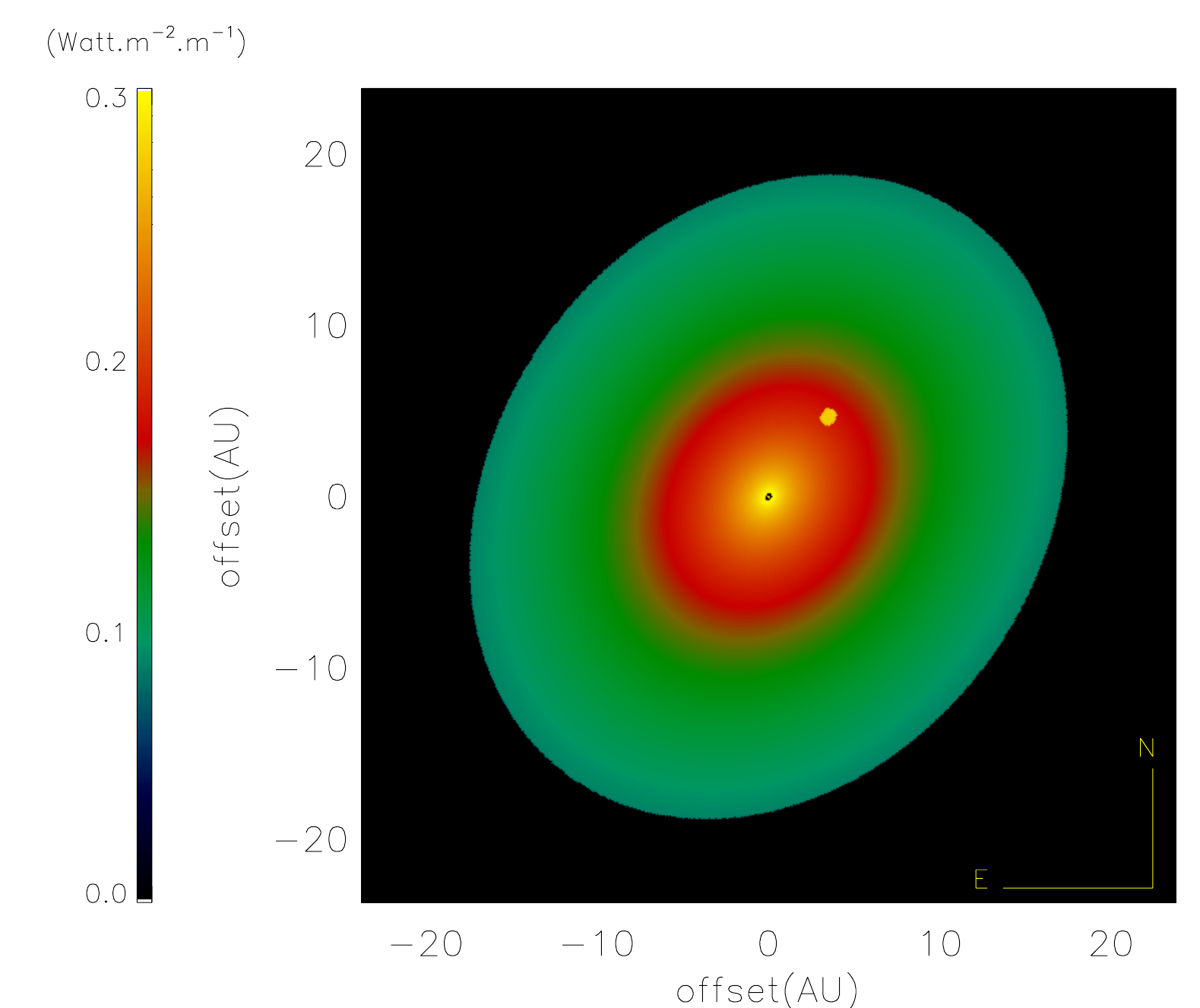}
\end{tabular}
\caption{One-component disk model with a bright feature. Top: The MIDI visibilities in 2007 and 2013. Middle-left: The Keck visibility. Middle-right: The SED from 2 $\mu$m to 20 $\mu$m. The black lines in whole plots represent observations and the blue ones represent the best model. Bottom: The synthetic image of MWC480 at $\lambda$=10 $\mu$m.}
 \label{hbalmeroii}
\end{figure*}

\subsection{Asymmetric model} 

With our axisymmetric model, we failed to reproduce the SED, the Keck and the MIDI visibilities simultaneously. For this reason, we consider in the following, the simplest situation that could produce azimuthal asymmetry, i.e., a bright feature in the disk.

 Therefore, from our one-component disk model, we added a bright feature along the direction of the projected baseline of the 2013 MIDI visibility. This choice is motivated by the fact that the clump should be invisible for the 2007 MIDI baseline (i.e. its projected distance onto the baseline direction is zero or close to zero). Our chosen azimuthal configuration for the clump enables that, without adding a temporal evolution of the azimuthal position of the clump in our model and therefore additional free parameters.
 
 Such a feature could be an embedded companion or a local dust concentration in a clump.
This bright feature or clump emission is modelled as an optically thick component having a black body emission $B_{\lambda}[T_{cl}(d_{cl})]$, where $T_{cl}$ is the temperature of the clump and $ d_{cl} $ is the distance of the clump from the central star. For each surface area element of the clump, the observer receives 
\begin{equation}
       dF_{\lambda,clump}=B_{\lambda} \left[  T_{\rm cl}(d_{\rm cl})\right]  \left(\frac{A_{\rm cl}}{D^{2}}\right)\,,
   \end{equation} 
   where $A_{\rm cl}$ is the elementary surface area of the clump that is defined by our pixel size in the simulated brightness maps of our model.
   
In this model, we set the best-fit parameter values obtained for the one-components disk model as explained in section 3.1. As mentioned above, the P.A. of the clump is set to the P.A. of the baseline used in 2013, i.e. $330^{\circ }$ (see Table 1) to minimise the parameters of the clump.  Therefore, the free parameters of the clump are following: 

\begin{enumerate}

\item[-] the distance of the clump ($d_{\rm cl}$),
\item[-] the size of the clump ($r_{\rm cl}$),
\item[-] the temperature of the clump ($T_{\rm cl}$).

\end{enumerate}

 In order to obtain the best-fit parameters, we carried out the same process as explained in section. 3.1 to reach the minimum $ \chi^{2} $. Fig. 7 illustrates the $ \chi^{2} $ maps for the best parameters. The value of the ${\chi^{2}}_{\rm r \hspace{2pt} \rm tot} $ for the wide range of parameters was 0.31.
Table 4 shows the wide and narrow range of the values that were scanned for each parameter. The uncertainties of the parameters were derived as in section 3.1. Fig. 8 presents the corresponding best-fit MIDI visibilities and the SED and the synthetic image for this best-fit model at 10 $\mu$m. 

The agreement with the 2013 MIDI visibility is now better. A partly resolved bright feature along the 2013 baseline direction allows to decrease the visibility towards longer wavelengths.  The agreement with the SED between 8 and 13 $\mu$m (silicate emission feature) is also better than in the axisymmetric case. All the other data are still consistently reproduced within their error bars.

In Fig. 9, we compared the photometry for the axisymmetric time variable models and the asymmetric model for the maximum grain size of 10 $\mu$m. We do show that the photometry changes in the time variable models is more significant than the changes between our best-fit axisymmetric model discussed in section 3.1 and the asymmetric one.
   \begin{figure}
\centering
\begin{tabular}{c}
\includegraphics[clip=,width=0.8\hsize]{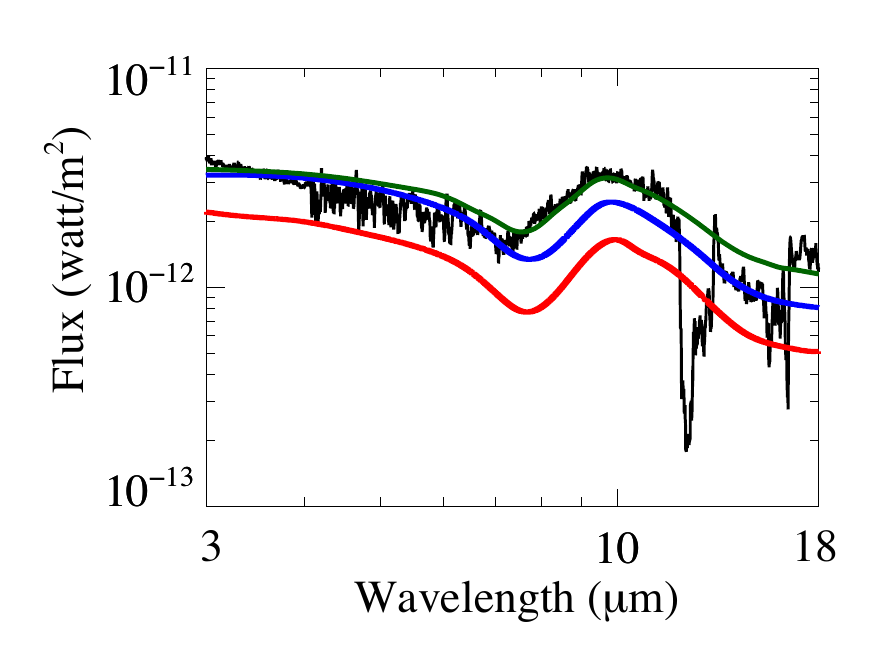} 
\end{tabular}
\caption{A comparison between the axisymmetric time variable models for 2007 and 2013 and the asymmetric model. The blue line is related to the best-fit model, which reproduces the MIDI visibility in 2007 alone, the red line is related to the model, which reproduces the 2013 MIDI visibility alone, and the green line is related to the asymmetric model.}
 \label{hbalmeroii}
\end{figure}


\section{Discussion}

Our simultaneous modeling of the SED as well as the mid-IR interferometric data of MWC480 favors the possibility of an asymmetric inner disk.
 Indeed, a one-component disk model including a bright clump in the inner disk gives better agreement with our data set.

This assumes that the azimuthal location of the bright feature was either similar in 2007 and 2013, or such that the clump was invisible for the 2007 baseline (i.e. its distance projected onto the baseline direction is zero). We may thus wonder if it is consistent for instance with a clump of dust in Keplerian rotation around the central star. Since the derived distance of the clump, i.e., $d_{\rm cl}$ = 5.45 $\pm$ 0.80 au is a projected one, we need to retrieve the physical distance, i.e. the unprojected distance, which is $d_{\rm u cl} $ = 5.5 $\pm$ 0.81 au. Thus, we use this latter distance to calculate the orbital period of the clump using the following equation:

\begin{equation}
       T=2 \pi    \sqrt{\frac{r^{3}}{G{M_{\star}}}},
   \end{equation} 

 where r is the unprojected distance of the clump ($d_{\rm u cl} $= 5.5 $\pm$ 0.81  au) and $M_{\star}$ is the mass of the central star (1.67 $\pm$ 0.07 M$\odot$ (Simon, Dutrey $\&$ Guilloteau, 2000)). With this parameters, the orbital period of this bright feature would be 9.9 $\pm$ 2.1 yrs, which is close to the time interval between the two MIDI observations, i.e., $\sim 7$ yrs.

The brightness ratio between the clump and its surrounding is nearly $ 6.5\times10^{2} $. This high brightness ratio results from the high optical depth of the clump compared to the very low optical depth of the disk at 5.45 au (the clump location of the model). The clump is assumed to be optically thick (see Eq. 13) while the disk optical depth is of the order of 4 $\times$ ${{10}^{-3}}$ at 10 $ \mu $m and varies from 2 $\times$ ${{10}^{-3}}$ to 3 $\times$ ${{10}^{-3}}$  in the range 8-13 $ \mu $m.

Binarity of the star can also cause an asymmetry in the disk. We tested this possibility with our models. In order to resolve the disk in the direction corresponding to the baseline orientation in 2013 and to decrease the visibility down to 0.6, we need a companion that is bright enough at the N band. However, such a stellar companion can be excluded since it would cause an additional NIR excess in the SED and a NIR visibility lower than the Keck one. Only a structure such as 'dusty clump' is cool enough  to contribute mainly at MIR and longer wavelengths without increasing too much the NIR emission.

A disk warping or spiral waves can also make asymmetries in the disk. For instance, a disk warping was used by Kraus et el. (2013) to explain the asymmetries in the inner disk of the star V1247 Orionis. Indeed, to reproduce the 2007 and 2013 MIDI data simultaneously, we need to add a clump of dust with a size of 0.85 au. Increasing this size will imply a visibility drop that is not seen in the flat 2007 MIDI visibility. Moreover, in order to explain both the visibility drop seen in the 2013 MIDI data and the flat shape of the 2007 MIDI visibility, one is bound to place the clump at a position angle that makes it invisible to the 2007 baseline (e.g., the same position angle as that of the 2013 MIDI data).
This shows that our results are very sensitive to the adopted size and position angle of the clump. To summarise, adding a spiral wave and /or a disk warping would allow to better resolve the disk and decrease the modelled visibilities below the observed ones and increase the level of the SED up the observed ones, since these structures are more extended radially and azimuthally.




 
\section{Summary and perspectives}

Using spectro-interferometry, we were able to resolve the circumstellar emission around the Herbig star MWC480. We performed a multi-wavelength modelling that aimed at reproducing the broadband SED, the Keck and the MIDI interferometric data. The modeling is based on a semi-analytical approach using a temperature and surface density-gradient one-component disk. 
Our aim is to constrain better the inner disk structure and in particular its axisymmetry, as suggested by previous interferometric studies. We concluded that: 

   \begin{enumerate}
    \item[-] an axisymmetric continuous model can not reproduce all our data, in particular the two MIDI measurements taken at perpendicular direction and the SED for the wavelengths between 8 to 18 $\mu$m.
     \item[-] Fitting independently the two MIDI visibilities implied photometric changes that are not consistent with the 2007 and 2013 IR spectra.

      \item[-] A better agreement is obtained
by considering in addition a bright feature that we determined to be located at 5.45 au from the star with a temperature of 400 K. 
If we assume that, in 2013, the clump is indeed aligned with the 2013 MIDI baseline, the fact that this clump is invisible to the 2007 baseline (i.e. its projected distance onto the direction of the 2007 baseline is zero) could be roughly explained in the frame of clump in Keplerian rotation.    

Reconstructing images of this disk with the upcoming second-generation VLTI instrument MATISSE (the Multi AperTure mid-Infrared SpectroScopic Experiment, Lopez et al. 2006) constitutes a unique perspective to further assess the nature of the inner region of the disk around MWC480. MATISSE will recombine up to four telescopes in the mid-IR (from 3 to 13~$\mu$m) and will thus provide a more complete UV coverage with different baseline orientations and closure phase measurements, that will be used to reveal unambiguously brightness asymmetries.  
 \end{enumerate}   
   
       \section{ACKNOWLEDGEMENT}
      The initial MIDI observation of this work has been obtained by Di Folco in 2007. We could again observe MWC480 in 2013. This work is supported at the Aerospace Corporation by the Independent Research and Development program. N.J. acknowledges the Ph.D. financial support from the Observatoire de la Cote d’Azur and the PACA Region. The authors wish to thank C. A. Grady, A. Crida, A. Meilland, P.Tanga and S. Casassus for useful exchanges and V. Girault for helping in the 2013 observations. 
    
\bibliographystyle{mn2e}
\bibliography{mn}

\end{document}